\documentclass[11pt]{article}

\usepackage[english]{babel}
\usepackage[latin1]{inputenc}
\usepackage{graphicx}
\usepackage{color}
\usepackage[hidelinks]{hyperref}
\usepackage{amsmath,amsfonts}
\usepackage{natbib}
\usepackage{bm}
\usepackage{enumerate}
\usepackage{multirow}
\usepackage{booktabs}
\usepackage{subcaption}
\usepackage[flushleft]{threeparttable}

\bibpunct{(}{)}{;}{a}{,}{,}

\title{\LARGE\textbf{Analysis of permanence time in emotional states: \\ A case study using educational software}} 

\author{\, \vspace{0.1cm} \\ \textbf{Helena Reis}\\University of S\~ao Paulo (USP), S\~ao Carlos, SP, Brazil \vspace{0.3cm} \\ \textbf{Danilo Alvares}\\Harvard T.H. Chan School of Public Health, Boston, MA, USA \vspace{0.3cm} \\ \textbf{Patricia Jaques}\\University of the Sinos Valley (UNISINOS), S\~ao Leopoldo, RS, Brazil \vspace{0.3cm} \\ \textbf{Seiji Isotani}\\University of S\~ao Paulo (USP), S\~ao Carlos, SP, Brazil}

\date{}

\begin{document} 

\baselineskip18pt

\maketitle 

\begin{abstract}
\noindent This article presents the results of an experiment in which we investigated how prior algebra knowledge and personality can influence the permanence time from the confusion state to frustration/boredom state in a computer learning environment. Our experimental results indicate that people with a neurotic personality and a low level of algebra knowledge can deal with confusion for less time and can easily feel frustrated/bored when there is no intervention. Our analysis also suggest that people with an extroversion personality and a low level of algebra knowledge are able to control confusion for longer, leading to later interventions. These findings support that it is possible to detect emotions in a less invasive way and without the need of physiological sensors or complex algorithms. Furthermore, obtained median times can be incorporated into computational regulation models (e.g. adaptive interfaces) to regulate students' emotion during the teaching-learning process.
\end{abstract}

\section{Introduction} \label{secintro}

Emotions have an important impact on learning, accelerating or hindering it \citep{pekrun2006,sullins2014,dmello20132}. Although most studies that investigate emotions in the educational context have focused on the basic emotions (e.g. sadness, anger, joy, or surprise), recent research provides evidence that non-basic emotions (e.g. engagement/flow, confusion, frustration, and boredom) are more frequent in computer-based learning (CBL) \citep{dmello2013}. 

Contrary to common sense, confusion is an emotion that should not be avoided in the learning context because it makes the students seek the knowledge and maintain focus and attention and is related to encouragement \citep{dmello20132}. Students commonly experience confusion in complex activities which occur throughout the entire school period. 
When confusion is detected and experienced, students need to engage in cognitive activities to solve their confusion.

Although confusion has been positively correlated with learning \citep{craig2004,dmello2007}, it should be regulated according to students' personality and prior knowledge to have an adequate duration \citep{dmello2007}. If confusion persists for a long time, it can become frustration or boredom \citep{graesser2011}. For instance, if a student has a neurotic personality (i.e., tends to have negative emotions) and he/she is a beginner in the subject or the task is complex, the confusion must be managed cautiously so as not to become boredom. However, one question remains: how specifically do students' personality and prior knowledge affect the permanence time in a confusion state?

To verify the relation of the permanence time in a confusion state with the students' personality and their prior knowledge on learning problems, we have developed an experiment with higher education students from three Brazilian public universities. This experiment was performed in more than 70 hours and involved 30 randomly selected students, 2 instructors and 2 coders, who analyzed the prior algebra knowledge and the students' personality. These students were also asked to solve algebra problems in a computer learning environment. The results obtained can be used to create less invasive and low cost alternative emotion detectors, unlike other types of detection, such as physiological sensors, which can be costly and make students uncomfortable \citep{shanabrook2012}. In addition, this approach supports emotional regulation models, in which the students' emotion can be regulated when they are feeling some negative emotion.

\section{Related works}

Emotion is a state constantly awakened and lived by individuals \citep{xiaolan2013}. Emotions can undergo several changes upon receiving a stimulus (i.e., a person can become angry, sad, or joyful). This change of emotion is called the transition state and can be influenced by initial emotion, emotional events, prior knowledge and individual personality characteristics \citep{gross2002}.

Studies \citep{graesser2005,dmello2014} have showed that, in CBL, emotions transit between engagement/flow, confusion and frustration/boredom. Confusion and engagement/flow have been positively correlated with learning, while frustration and boredom have been negatively correlated. Ideally, emotions should be regulated considering students' personality and knowledge, and should have a certain duration \citep{dmello2007}. For instance, academic risk theory contrasts adventurous students, who want to be challenged with difficult tasks, take the risk of failure, and manage negative emotions when it occurs, with students who take less risks, avoiding complex tasks and effectively minimize learning situations in which they are likely to fail and experience negative emotions \citep{clifford1988,dweck2006,meyer2006}.

In addition, it is necessary to identify who could benefit from an inductive intervention for a particular emotion \citep{craig2004,dmello2007}. Of course, confusing a beginner student or inducing confusion during high-risk learning activities is not a sensible strategy. Nowadays, these interventions are ideally suited for gifted students who get bored and lose interest in activities without challenges \citep{pekrun2010,dmello2014}. \citep{dmello2012}  investigated the emotional transitions\footnote{As students can feel more than one emotion each time, in this paper we are considering the dominant emotion in a moment in time and the transition to another dominant emotion \citep{larsen2001can} during the teaching-learning process.}, during the teaching-learning process (Table \ref{tab:transicoes}). Their results show evidence that confusion can lead to two other emotions: engagement/flow and frustration. 

\begin{table}[htb!]
\centering
\caption{Expected affective transitions (Adapted from \citep{dmello2012}).}
\label{tab:transicoes}
\begin{tabular}{|c|c|c|c|c|}
\hline
\multirow{2}{*}{Time $t_{i}$} & \multicolumn{4}{c|}{Time $t_{i}+1$}                       \\ \cline{2-5} 
                                             & ~Boredom~  & ~Flow~ & ~Confusion~ & ~Frustration~ \\ \hline
Boredom                                      &                & -             & -                 & ?    \\ \hline
Flow                                         & -              &               & +                 & -    \\ \hline
Confusion                                    & -              & +             &                   & +    \\ \hline
~Frustration~                                & +              & -             & ?                 &      \\ \hline
\end{tabular}
\end{table}
\vspace{-0.3cm}
\begin{table}[htb!]
\centering
\label{my-label}
\begin{tabular}{l}
\small (+) indicates that the transaction is expected. \\
\small (-) indicates that the transaction is highly unlikely. \\
\small (?) indicates that there is no explicit relation in the model.
\end{tabular}
\end{table}

When confusion is not handled appropriately (i.e., when instructors do not monitor the duration of confusion or the students' personality - tendency to emotional states) the student can become frustrated and then bored. When a student experiences negative emotions, such as frustration and boredom, he/she tends to remain in this state and not transit to positive emotional states, such as engagement/flow. On the other hand, students in flow state tend to remain engaged or transit alternately to confusion, which is positively correlated with learning \citep{dmello2007}. In a more general context, the transition from an  emotional state to another one can be modeled through a survival or reliability analysis, where the transition probabilities are obtained for each specified time \citep{meeker1998}.

\section{Method}

During the teaching-learning process, the student can experience several emotions (engagement/flow, confusion, frustration/boredom, etc.) and these emotions can transit from one to another. The change from an emotional state to another one can depend on several factors, including personality and prior knowledge on the subject. For instance, a beginner student with personality tending to negative emotions (e.g., neuroticism) can easily move from confusion to frustration/boredom when he/she strives to solve a problem or has a long  period of confusion.
 
This section describes the design and planning used in our experiment to investigate whether the personality traits (neuroticism and extroversion) and algebra knowledge of students affect their permanence times from the confusion state to frustration/boredom state during the use of a computer learning environment.

\subsection{Research questions}

This work aims to answer the following research questions (RQ):

\textbf{RQ}$_{1}$: Do personality traits influence the permanence time from confusion to frustration/boredom in an educational software?

\textbf{RQ}$_{2}$: Does algebra proficiency influence the permanence time from confusion to frustration/boredom in an educational software?

\textbf{RQ}$_{3}$: What is the average permanence time spent by a student (with different combinations of personality traits and algebra proficiency) from confusion to frustration/boredom in an educational software?

\subsection{Participants}

We gathered information from 30 randomly selected students. 13\% are women (corresponding to 4 people) aged 21-22 years (mean age of 21.5 years) and 87\% are men (corresponding to 26 people) aged 19-34 years (mean age of 26.5 years). All participants were invited through direct contact and are undergraduate students in areas related to Computer Science and Software Engineering, except 4 male participants who are undergraduate students in Industrial Design (2 people), Production Engineering (1 person) and Geography (1 person).

\subsection{Materials}

For the execution of the experiment, the following instruments were used: (i) questionnaire with personal questions, (ii) multiple-choice test, (iii) personality trait scale and (iv) equation solving test. The personal questionnaire aims to know the profile of participants and contains questions about personal data, such as whether the participants have already used some educational software before and their prior knowledge on algebra. The algebra test covered five multiple choice questions which involved first and second degree equations, determinants, factorials, and logarithms. These questions were suggested by two math teachers and they were separated into three difficult levels: basic (2), intermediate (2) and difficult (1). Each correct question was assigned the value 0.2 points, so the total points for each student can be 0, 0.2, 0.4, 0.6, 0.8 or 1. 

The personality trait scale\footnote{Available at \url{https://personalitatem.ufs.br/inventory/home.xhtml}.} assessed the participant's personality for neuroticism and extroversion indexes. Each of these indexes varies from 0 to 1, where 1 indicates greater presence of characteristic summarized by the index. This test was based on the five-factor model and is written in Portuguese. 
Finally, nine algebra questions were proposed in an educational software\footnote{Available at \url{http://acubo.tecnologia.ws/aluno.html}.}, all in the same scope previously tested. The nine questions also involved first and second degree equations, determinants, factorials, and logarithms. They were suggested by two math teachers and they were separated into three difficult levels: basic (3), intermediate (4) and difficult (2).

\subsection{Procedure}

\begin{figure}[htb!]
\centering
\includegraphics[width=8cm]{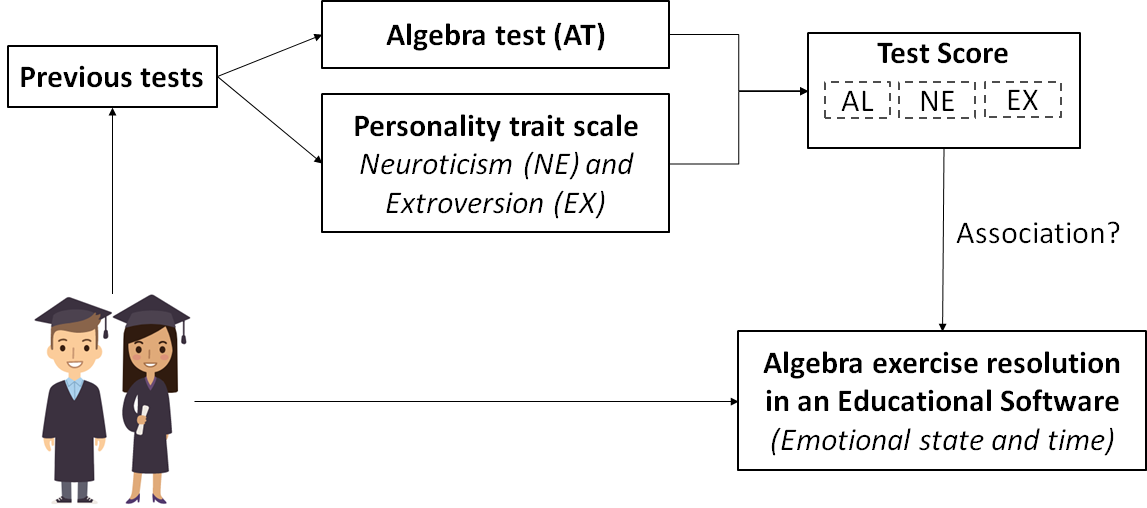}
\caption{Procedure of the experiment.}
\label{fig:mesh1}
\end{figure}

Each participant had an hour and a half to perform the experiment (Figure \ref{fig:mesh1}). First, students were asked to fill in the personal questionnaire in 10 minutes. Afterwards, students answered the multiple choice test and then the personality trait scale, which had a total duration of 30 minutes. After this initial phase, students accessed a system for solving algebra problems. First, they should fill out information with their personal data. Hence, they solved nine algebra problems (scratch papers were provided) and entered their final answer in the system. The students' face were recorded while they were solving the equations for later analyses of their emotions (confusion, frustration and boredom).

As previously mentioned, the analysis of the video was performed by two coders,  one graduates from
Computer Science and one from Industrial Design. According to \citep{sebe2005}, the recognition of facial expressions by humans has an accuracy of approximately 87\%, making it possible that people with no training in Psychology and without any tool to measure emotions can recognize different types of emotions by the face.

The coders annotated the emotions students experienced and the permanence time in each of them during problem solving. They separately annotated the beginning and ending time that each student expressed, by face, the emotions engagement/flow, confusion, frustration, or boredom. Then, they discussed together the annotations and reached an agreement. The coding of the facial expressions was performed according to guidelines suggested by \citep{lera2007}, in which they propose 10 heuristics of human behavior in order to infer what emotions humans are experiencing at a given moment.

\section{Statistical Modeling} \label{secmodel}

This study aims to determine the permanence time from a confusion state to a frustration/boredom state for each student. So we used a statistical model that describes the permanence time in a state until an event of interest occurs. This type of approach is known as survival or reliability analysis, and its main objective is to know the behavior of a given population as to the time of occurrence of one or more events of interest \citep{klein2012}. We have opted for a Bayesian inferential analysis, in which all unknown quantities (e.g., model parameters) can be modeled by means of probability distributions \citep{bernardo1994}.

Our modeling for the permanence time from confusion to frustration/boredom for $i^{th}$ student, $i=1,\ldots,30$, will be characterized by a Weibull proportional hazards model with fragility \citep{sahu1997}, given by:
\begin{equation}
h_{i}(t \mid {\boldsymbol \theta}, w_{i}, {\bf x}_{i}) = \lambda\,\alpha\,t^{\alpha-1}\,\mbox{exp}\Big(\beta_{1}\,x_{1i}+\beta_{2}\,x_{2i}+\beta_{3}\,x_{3i}\Big) \, w_{i}, \label{model}
\end{equation}

\noindent where $h_{i}(t \mid \cdot)$ is a risk function for student $i$ at time $t$. The parameters $\beta_{1}$, $\beta_{2}$ and $\beta_{3}$ are the coefficients associated with students scores $i$ in the preliminary algebra test ($x_{1i}$), neuroticism index ($x_{2i}$) and extroversion index ($x_{3i}$), respectively. $\lambda$ and $\alpha$ are scale and shape parameters of the Weibull distribution that defines the baseline risk function described by $\lambda\,\alpha\,t^{\alpha-1}$. The frailty (or random effect) for the student $i$ is described by $w_{i} \sim \mbox{Gamma}(\eta,\eta)$, where its variance is given by $\kappa=1/\eta$. The parameter and variable vectors are defined as ${\boldsymbol \theta}=(\beta_{1}, \beta_{2}, \beta_{3}, \lambda, \alpha, \eta)^{\top}$ and ${\bf x}_{i}=(x_{1i}, x_{2i}, x_{3i})^{\top}$.

We assume prior independence as a default specification. In addition, we elicit vague proper marginal prior distributions, in order to give all inferential prominence to the data:
\begin{align}
		\begin{array}{c} 
					\pi(\beta_{1}) = \pi(\beta_{2}) = \pi(\beta_{3}) = \pi(\log(\lambda))  =  \mbox{Normal}(0,1000), \\
					\pi(\alpha) = \pi(\eta)  =  \mbox{Gamma}(0.01,0.01).
		\end{array}
		\label{previas}
\end{align}

The posterior distribution $\pi({\boldsymbol \theta} \mid \mathcal{D})$ is not obtained analytically ($\mathcal{D}$ represents the collected data), so we approximate it using Markov chain Monte Carlo (MCMC) \citep{gamerman2006} with the \texttt{WinBUGS} software \citep{lunn2000}.

\section{Discussion of results}

In this paper, we aim to study the permanence time of students from confusion to frustration/boredom, given their algebra knowledge and personality. The presented results below are from the model (\ref{model}) with the marginal prior distributions defined as in (\ref{previas}), where we use the following MCMC configuration: 3 Markov chains with 500000 iterations (after burning of 50000) and storage every 500 iterations to reduce autocorrelation in the posterior sample.

Table \ref{table-results1} summarizes the posterior estimates of the parameters $\beta_{1}$, $\beta_{2}$ and $\beta_{3}$ of the model (\ref{model}) with prior distributions (\ref{previas}).

\begin{table}[htb]
\centering
\caption{Posterior summary of the parameters of interest for the time from confusion to frustration/boredom.}
\begin{tabular}{|c|ccccc|}
\hline
Parameter      &~~Mean~~ & ~~SD~~ & ~~$2.5\%$~~ & ~~$50\%$~~  & ~~$97.5\%$~~ \\
\hline
$\beta_{1}$    & -1.970  & 0.672  & -3.334  & -1.956  &  -0.725 \\
$\beta_{2}$    & ~0.721  & 0.634  & -0.602  & ~0.725  &  ~1.943 \\
$\beta_{3}$    & -0.828  & 0.710  & -2.202  & -0.835  &  ~0.581 \\
\hline
\end{tabular}
\label{table-results1}
\end{table}

The interpretation of the results from the Bayesian approach is simple and fundamentally based on quantities of interest, such as mean, standard deviation (SD) and quartiles, from probability distributions, called posterior (or \textit{a posteriori}) distributions. In addition, the interpretation of mean signal of each parameter is counter-intuitive, because in the case of negative signal, the higher the value of the variable referring to this parameter, the longer the time until the student experiences the event of interest. 

As for the question \textbf{RQ}$_{2}$, based on presented results in Table~\ref{table-results1}, we have evidences that the more algebra knowledge, the longer the time until the student in the confusion state becomes frustrated/bored with the exercise, i.e., there is a positive association. The answer to \textbf{RQ}$_{1}$ is divided into two parts, where the first one refers to extroversion index and the second one to neuroticism index. The interpretation for extroversion index is analogous to the algebra knowledge, since the higher the extroversion index, the longer the permanence time between confusion and frustration/boredom. On the other hand, an increase in the neuroticism index leads to a reduction in the permanence time from confusion to frustration/boredom, i.e., the student gives up on the problem solving more quickly (negative association).

From the posterior distribution of ${\boldsymbol \theta}$, we can calculate derived quantities that help us to answer \textbf{RQ}$_{3}$, such as the median time of permanence time from confusion to frustration/boredom. This median time $T$ is obtained when the survival function $S_{i}(T)$ for a student $i$ takes the value 0.5 and is given by:
\begin{equation}
T = \left[\frac{-\log(0.5)}{\lambda\,\mbox{exp}\Big(\beta_{1}\,x_{1i}+\beta_{2}\,x_{2i}+\beta_{3}\,x_{3i}\Big) \, w_{i}}\right]^{1/\alpha}. \label{mediantime}
\end{equation}

It is worth mentioning that, in survival analysis, right-censored data make the median more informative than the mean, i.e., participants who do not experience the event of interest - frustration/boredom state - during the study time would take the average to higher values. In Bayesian terms, we can calculate the posterior mean of the median time (\ref{mediantime}) for a generic individual $i$, given his/her variables ${\bf x}_{i}$, by the following equation:
\begin{align}
\mbox{E}\big[S_{i}(T \mid {\boldsymbol \theta}, {\bf x}_{i}) \mid \mathcal{D}\big] & = \int S_{i}(T \mid {\boldsymbol \theta}, w_{i}, {\bf x}_{i}) \, \pi(w_{i},{\boldsymbol \theta} \mid \mathcal{D}) \, \mbox{d}(w_{i},{\boldsymbol \theta}) \nonumber \\
   & \approx \frac{1}{K}\sum_{k=1}^{K} S_{i}(T \mid {\boldsymbol \theta}^{(k)}, w_{i}^{(k)}, {\bf x}_{i}), \label{posteriormediantime}
\end{align}
\noindent where ${\boldsymbol \theta}^{(k)}$ and $w_{i}^{(k)}$ are $k^{th}$ values of posterior sample $\pi(w_{i},{\boldsymbol \theta} \mid \mathcal{D})$. The approximation (\ref{posteriormediantime}) is carried out by Monte Carlo integration \citep{niederreiter2003}.

To exemplify the obtained results and answer \textbf{RQ}$_{3}$, Table \ref{table-perfis1} shows the posterior mean of median time of the permanence time from confusion to frustration/boredom (\ref{posteriormediantime}) with different configurations of prior algebra knowledge and scores of neuroticism and extroversion indexes.

\begin{table}[htb!]
\centering
\caption{Posterior mean of median time of the permanence time from confusion to frustration/boredom for different profiles.}
\begin{tabular}{|c|ccccccccc|}
\hline
~Variable~    &  \multicolumn{9}{c|}{Profile} \\
\hline
$x_{1}$     & 0   & 0.5  & 1   & 0   & 0   & 1   & 1   & 0   & 1 \\
$x_{2}$     & 0   & 0.5  & 0   & 1   & 0   & 1   & 0   & 1   & 1 \\
$x_{3}$     & 0   & 0.5  & 0   & 0   & 1   & 0   & 1   & 1   & 1 \\
\hline
Time & ~34~  & ~76~   & ~159~ & ~19~  & ~64~  & ~90~  & ~313~ & ~37~  & ~175~ \\
\hline
\end{tabular}
\label{table-perfis1}
\end{table}

From the results, the influence of each variable at median time of the permanence time from confusion to frustration/boredom is evident. For instance, a student with a ``median'' configuration (i.e., 0.5 for all variables) would take on average 76 seconds to migrate between the states of interest. We also can note that students with a maximum score (value 1) for the prior algebra test and the extroversion index, and a minimum score (value 0) for the neuroticism index would require, on average, 313 seconds to pass from confusion to frustration/boredom. On the other hand, when the student has a maximum score for the neuroticism index and a minimum for the prior algebra test and the extroversion index, he/she takes, on average, 19 seconds.Note that we did not include in the model (1) a covariate describing the difficulty level of algebra questions. This is due to the fact that, in our preliminary analyzes, this covariate did not present relevant differences between the three difficult levels (basic, intermediate and difficult).

\section{Threats to validity}

A possible threat to validity of the results is the representativeness of the sample, since all individuals who participated in the study are undergraduate students. In this way, it is not possible to generalize the results to the entire student population. From the statistics point of view, this problem can be circumvented with repetitions of this study in different samples of undergraduate students. Although there was concern in assessing the permanence time from confusion to frustration/boredom, the usability of the software for the algebra test may have prevented some students to complete the exercises. Another threat to be considered is the use of two people to code students emotions by face observation, leading to an interpersonal bias as to the accuracy in the permanence time.

\section{Conclusions}

We aimed to study the permanence time of students from a confusion state to a frustration/boredom state, given their algebra knowledge and personality. The results of our experiment suggest that prior algebra knowledge and personality traits affect the permanence time from confusion to frustration/boredom during the learning process. Notably, students with a high neuroticism index and low score in the algebra test cannot deal very well with the confusion, remaining less time in this emotion compared to students with high extroversion index and low score in the algebra test. This means that neurotic students who are also beginners in algebra spend less time confused and feel frustration/boredom more quickly compared to extroverted student with the same level of algebra knowledge.

We believe that these preliminary results can help in the elaboration of computational models of emotional regulation of students. The permanence time in the confusion state can be integrated with other information, for instance, physiological sensors or automatic facial expressions. This information can contribute to emotion control using interfaces, which detect personality traits and the beginning of the confusion state, adapting elements for beginner students with little tolerance in the permanence time of confusion. In addition, as future work, the results may help with the investigation of student self-efficacy. Other benefit of this experiment was the provision of a replication package\footnote{\url{http://goo.gl/YtGn7H}}, which can be used by other researchers for the same purpose. 

Focusing on the statistical approach, the Bayesian perspective for survival analysis was of paramount importance, since we had a small data set and wanted to interpret derived quantities based on model parameters (questions \textbf{RQ}$_{1}$, \textbf{RQ}$_{2}$ and \textbf{RQ}$_{3}$). Furthermore, the estimated permanence time between confusion and frustration/boredom for any new student profile is easily calculated, providing quick decision-making with respect to emotional regulation.

\bibliographystyle{plainnat}
\bibliography{references}

\begin{thebibliography}{27}
\providecommand{\natexlab}[1]{#1}
\providecommand{\url}[1]{\texttt{#1}}
\expandafter\ifx\csname urlstyle\endcsname\relax
  \providecommand{\doi}[1]{doi: #1}\else
  \providecommand{\doi}{doi: \begingroup \urlstyle{rm}\Url}\fi

\bibitem[Bernardo and Smith(1994)]{bernardo1994}
J.~M. Bernardo and A.~F.~M. Smith.
\newblock \emph{Bayesian theory}.
\newblock John Wiley \& Sons, 1st edition, 1994.

\bibitem[Clifford(1988)]{clifford1988}
M.~M. Clifford.
\newblock Failure tolerance and academic risk-taking in ten- to twelve-year-old
  students.
\newblock \emph{British Journal of Educational Psychology}, 58\penalty0
  (1):\penalty0 15--27, 1988.

\bibitem[Craig et~al.(2004)Craig, Graesser, Sullins, and Gholson]{craig2004}
S.~Craig, A.~Graesser, J.~Sullins, and B.~Gholson.
\newblock Affect and learning: {A}n exploratory look into the role of affect in
  learning with {AutoTutor}.
\newblock \emph{Journal of Educational Media}, 29\penalty0 (3):\penalty0
  241--250, 2004.

\bibitem[D'Mello(2012)]{dmello2012}
S.~D'Mello.
\newblock \emph{Encyclopedia of the sciences of learning}, chapter Monitoring
  affective trajectories during complex learning, pages 2325--2328.
\newblock Springer US, 2012.

\bibitem[D'Mello and Calvo(2013)]{dmello2013}
S.~D'Mello and R.~A. Calvo.
\newblock Beyond the basic emotions: {W}hat should affective computing compute?
\newblock In \emph{CHI '13 Extended Abstracts on Human Factors in Computing
  Systems}, pages 2287--2294, 2013.

\bibitem[D'Mello and Graesser(2013)]{dmello20132}
S.~D'Mello and A.~Graesser.
\newblock {AutoTutor and Affective Autotutor}: {L}earning by talking with
  cognitively and emotionally intelligent computers that talk back.
\newblock \emph{ACM Transactions on Interactive Intelligent Systems},
  2\penalty0 (4):\penalty0 23:1--23:39, 2013.

\bibitem[D'Mello et~al.(2007)D'Mello, Picard, and Graesser]{dmello2007}
S.~D'Mello, R.~W. Picard, and A.~Graesser.
\newblock Toward an affect-sensitive {AutoTutor}.
\newblock \emph{IEEE Intelligent Systems}, 22\penalty0 (4):\penalty0 53--61,
  2007.

\bibitem[D'Mello et~al.(2014)D'Mello, Lehman, Pekrun, and Graesser]{dmello2014}
S.~D'Mello, B.~Lehman, R.~Pekrun, and A.~Graesser.
\newblock Confusion can be beneficial for learning.
\newblock \emph{Learning and Instruction}, 29:\penalty0 153--170, 2014.

\bibitem[Dweck(2006)]{dweck2006}
C.~S. Dweck.
\newblock \emph{Mindset: {T}he new psychology of success}.
\newblock Random House Incorporated, 1st edition, 2006.

\bibitem[Gamerman and Lopes(2006)]{gamerman2006}
D.~Gamerman and H.~F. Lopes.
\newblock \emph{Markov chain {M}onte {C}arlo: {S}tochastic simulation for
  {B}ayesian inference}.
\newblock Chapman \& Hall/CRC, 2nd edition, 2006.

\bibitem[Graesser and D'Mello(2011)]{graesser2011}
A.~Graesser and S.~K. D'Mello.
\newblock \emph{New perspectives on affect and learning technologies}, chapter
  Theoretical perspectives on affect and deep learning, pages 11--21.
\newblock Springer, 2011.

\bibitem[Graesser et~al.(2005)Graesser, Chipman, Haynes, and
  Olney]{graesser2005}
A.~C. Graesser, P.~Chipman, B.~C. Haynes, and A.~Olney.
\newblock {AutoTutor}: {A}n intelligent tutoring system with mixed-initiative
  dialogue.
\newblock \emph{IEEE Transactions on Education}, 48\penalty0 (4):\penalty0
  612--618, 2005.

\bibitem[Gross(2002)]{gross2002}
J.~J. Gross.
\newblock Emotion regulation: {A}ffective, cognitive, and social consequences.
\newblock \emph{Psychophysiology}, 39\penalty0 (3):\penalty0 281--291, 2002.

\bibitem[Kleinbaum and Klein(2012)]{klein2012}
D.~Kleinbaum and M.~Klein.
\newblock \emph{Survival analysis: {A} self-learning text}.
\newblock Springer-Verlag, 3th edition, 2012.

\bibitem[Larsen et~al.(2001)Larsen, McGraw, and Cacioppo]{larsen2001can}
Jeff~T Larsen, A~Peter McGraw, and John~T Cacioppo.
\newblock Can people feel happy and sad at the same time?
\newblock \emph{Journal of personality and social psychology}, 81\penalty0
  (4):\penalty0 684, 2001.

\bibitem[Lera and Garreta-Domingo(2007)]{lera2007}
E.~Lera and M.~Garreta-Domingo.
\newblock Ten emotion heuristics: {G}uidelines for assessing the user's
  affective dimension easily and cost-effectively.
\newblock In \emph{Proceedings of the 21st BCS HCI Group Conference}, volume~2,
  pages 163--166, 2007.

\bibitem[Lunn et~al.(2000)Lunn, Thomas, Best, and Spiegelhalter]{lunn2000}
D.~J. Lunn, A.~Thomas, N.~Best, and D.~Spiegelhalter.
\newblock Winbugs - {A} {B}ayesian modelling framework: concepts, structure,
  and extensibility.
\newblock \emph{Statistics and Computing}, 10\penalty0 (4):\penalty0 325--337,
  2000.

\bibitem[Meeker and Escobar(1998)]{meeker1998}
W.~Q. Meeker and L.~A. Escobar.
\newblock \emph{Statistical methods for reliability data}.
\newblock Wiley-Interscience, 1st edition, 1998.

\bibitem[Meyer and Turner(2006)]{meyer2006}
D.~K. Meyer and J.~C. Turner.
\newblock Re-conceptualizing emotion and motivation to learn in classroom
  contexts.
\newblock \emph{Educational Psychology Review}, 18\penalty0 (4):\penalty0
  377--390, 2006.

\bibitem[Niederreiter(2003)]{niederreiter2003}
H.~Niederreiter.
\newblock Some current issues in {quasi-Monte Carlo methods}.
\newblock \emph{Journal of Complexity}, 19\penalty0 (3):\penalty0 428--433,
  2003.

\bibitem[Pekrun(2006)]{pekrun2006}
R.~Pekrun.
\newblock The control-value theory of achievement emotions: {A}ssumptions,
  corollaries, and implications for educational research and practice.
\newblock \emph{Educational Psychology Review}, 18\penalty0 (4):\penalty0
  315--341, 2006.

\bibitem[Pekrun et~al.(2010)Pekrun, G{\"o}tz, Daniels, Stupnisky, and
  Perry]{pekrun2010}
R.~Pekrun, T.~G{\"o}tz, L.~M. Daniels, R.~H. Stupnisky, and R.~P. Perry.
\newblock Boredom in achievement settings: {E}xploring control-value
  antecedents and performance outcomes of a neglected emotion.
\newblock \emph{Journal of Educational Psychology}, 102\penalty0 (3):\penalty0
  531--549, 2010.

\bibitem[Sahu et~al.(1997)Sahu, Dey, Aslanidou, and Sinha]{sahu1997}
S.~K. Sahu, D.~K. Dey, H.~Aslanidou, and D.~Sinha.
\newblock A {W}eibull regression model with gamma frailties for multivariate
  survival data.
\newblock \emph{Lifetime Data Analysis}, 3\penalty0 (2):\penalty0 123--137,
  1997.

\bibitem[Sebe et~al.(2005)Sebe, Cohen, Gevers, and Huang]{sebe2005}
N.~Sebe, I.~Cohen, T.~Gevers, and T.~S. Huang.
\newblock Multimodal approaches for emotion recognition: {A} survey.
\newblock In \emph{Proceedings of the SPIE}, volume 5670, pages 56--67, 2005.

\bibitem[Shanabrook et~al.(2012)Shanabrook, Arroyo, and Woolf]{shanabrook2012}
D.~H. Shanabrook, I.~Arroyo, and B.~P. Woolf.
\newblock Using touch as a predictor of effort: {W}hat the ipad can tell us
  about user affective state.
\newblock In \emph{User Modeling, Adaptation, and Personalization}, volume
  7379, pages 322--327, 2012.

\bibitem[Sullins and Graesser(2014)]{sullins2014}
J.~Sullins and A.~C. Graesser.
\newblock The relationship between cognitive disequilibrium, emotions and
  individual differences on student question generation.
\newblock \emph{International Journal of Learning Technology}, 9\penalty0
  (3):\penalty0 221--247, 2014.

\bibitem[Xiaolan et~al.(2013)Xiaolan, Lun, Xin, and Zhiliang]{xiaolan2013}
P.~Xiaolan, X.~Lun, L.~Xin, and W.~Zhiliang.
\newblock Emotional state transition model based on stimulus and personality
  characteristics.
\newblock \emph{China Communications}, 10\penalty0 (6):\penalty0 146--155,
  2013.

\end{thebibliography}

\end{document}